# Pipeline for Advanced Contrast Enhancement (PACE) of chest X-ray in evaluating COVID-19 patients by combining bidimensional empirical mode decomposition and CLAHE

G. Siracusano,[1,*] A. La Corte,[1] M. Gaeta,[2,*] G. Cicero,[2] M. Chiappini,[3,4] G. Finocchio[5,*]

[1]Department of Electric, Electronic and Computer Engineering, University of Catania, Viale Andrea Doria 6, 95125 Catania, Italy

[2]Department of Biomedical Sciences, Dental and of Morphological and Functional Images, University of Messina, Via Consolare Valeria 1, 98125 Messina, Italy

[3]Istituto Nazionale di Geofisica e Vulcanologia (INGV), Via di Vigna Murata 605, I-00143 Roma, Italy

[4]Maris Scarl, via Vigna Murata 606, 00143, Roma, Italy

[5]Department of Mathematical and Computer Sciences, Physical Sciences and Earth Sciences, V.le F. Stagno D'Alcontres 31, 98166, University of Messina, Italy

*giulio.siracusano@gmail.com, mgaeta@unime.it, gfinocchio@unime.it

**Abstract.**
COVID-19 is a new pulmonary disease which is driving stress to the hospitals due to the large number of cases worldwide. Imaging of lungs can play a key role in monitoring of the healthy status. Non-contrast chest computed tomography (CT) has been used for this purpose, mainly in China, with a significant success. However, this approach cannot be used massively mainly for both high risk and cost and in some countries also because this tool is not extensively available. Alternatively, chest X-ray, although less sensitive than CT-scan, can provide important information about the evolution of pulmonary involvement during the disease, this aspect is very important to verify the response of a patient to treatments. Here, we show how to improve the sensitivity of chest X-ray via a nonlinear post processing tool, named PACE, combining properly fast and adaptive bidimensional empirical mode decomposition and contrast limited adaptive histogram equalization (CLAHE). The results show an enhancement of the image contrast as confirmed by three widely used metrics: (*i*) contrast improvement index, (*ii*) entropy, and *(iii)* measure of enhancement. This improvement gives rise to a detectability of more lung lesions as identified by two radiologists, which evaluate the images separately, and confirmed by CT-scans. Based on our findings this method is proved as a flexible and effective way for medical image enhancement and can be used as a post-processing step for medical image understanding and analysis.

## I. Introduction

Non-contrast chest Computed Tomography (CT) has many advantages over the planar chest X-ray which include the better spatial and densitometric resolution and the possibility of more clear identification of morphologic features of lesions.[1] Although chest CT can provide early diagnosis of COVID-19, it is not sufficient for that purpose considering that is aspecific, then for the World Health Organization the reference standard in the diagnosis of COVID-19 is the transcription-polymerase chain reaction which detects viral nucleic acid.[2] As already discussed in literature, chest CT can be used for the study of COVID-19 progression in patients providing quantitative information.[3] However, this approach needs several scans in a short window time. This fact gives rise to additional risks including increased possibility of cancer induction, because CT scan can expose a patient to as much radiations as 20-70 chest X-rays (CXR).[4] In addition, when used for evaluating COVID-19 patients, difficult and time consuming procedures of decontamination have to be set after each scan originating either increasing cost for patient and reducing availability time of the CT. Recent guidelines of North American Radiology Scientific Expert Panel have assessed that *portable* CXR has to be considered as the main imaging approach in evaluating COVID-19 patients[5] which not only reduces radiations for patients, but also avoids their transportation. The additional advantage of portable CXR is the possibility to monitor patients in the intensive care units (ICUs) which are more than 5% of the total known cases of COVID-19. Recent clinical studies [6,7] confirm that CXRs can be used in describing some features in COVID-19 patients. The protocol to use CXR for evaluating COVID-19 patients has been set in the Hospital 'Policlinico G. Martino' of Messina (Italy) since the beginning of February 2020. In particular, an anteroposterior grey scale CXR in supine position has been acquired for each patient at bed with a portable X-ray equipment. Those images are often of low quality for the environment difficulties [8–10] and for non-collaborative and severely ill patients in most cases [11,12] causing many different artifacts originating inhomogeneities in luminance distribution of radiograms.[13] To overcome these limitations that can impact on diagnostic effectiveness, here we develop a nonlinear post processing tool, we named Pipeline for Advanced Contrast Enhancement (PACE) which is aimed at improving the image quality of the CXR as evaluated in terms of contrast improvement index (CII), image entropy (ENT), and the measurement of enhancement (EME).

Fig. 1 shows a block diagram of the proposed Pipeline for Advanced Contrast Enhancement (PACE). Firstly, the CXR is processed with the Fast and Adaptive Bidimensional Empirical Mode Decomposition (FABEMD) which decomposes the input image into multiple hierarchical components known as bidimensional intrinsic mode functions (BIMFs) and a bidimensional residual image (BR), based on the local spatial variations or scales of the image. The residual image is then filtered with a Homomorphic Filter (HMF) where the kernel function is represented by a High-Frequency Emphasis Filter (HEF). The CXR is then reconstructed by recombining BIMFs with the filtered residual image and the Contrast Limited Adaptive Histogram Equalization (CLAHE) is finally applied to improve the contrast and generate an enhanced CXR (ECXR) image. The performance of the algorithm has been evaluated in 79 patients showing a stable and remarkable improvement on each of the considered metrics against state-of-the-art methods.

To evaluate the clinical impact of the post processed images, two radiologists have independently analyzed the lung lesions. They have subjectively found in all the enhanced images a clearer definition of the lesions already detectable in the original images and interestingly more lesions have been found in 8 cases (>10%). In 3 out of 8 cases the baseline CXR have been evaluated as negative. The positivity in the enhanced images of those cases has been confirmed by chest CT. This work paves the way for the development of sophisticated post processing tools to improve the image quality of the cost-effective portable CXR for the monitoring of COVID-19 patients and all patients in intensive critical unit as well. We argue this research can be also used as a support of clinical activities either in poor regions where CT is not available and hospitals in developed countries at the peak of pandemic COVID-19. In addition, PACE can be used as a tool for preprocessing the data or data augmentation in machine/deep learning approaches with a possible application in COVID-19 detection considering, current efforts devoted to develop deep learning model for early diagnosis of COVID-19. [14,15,16]

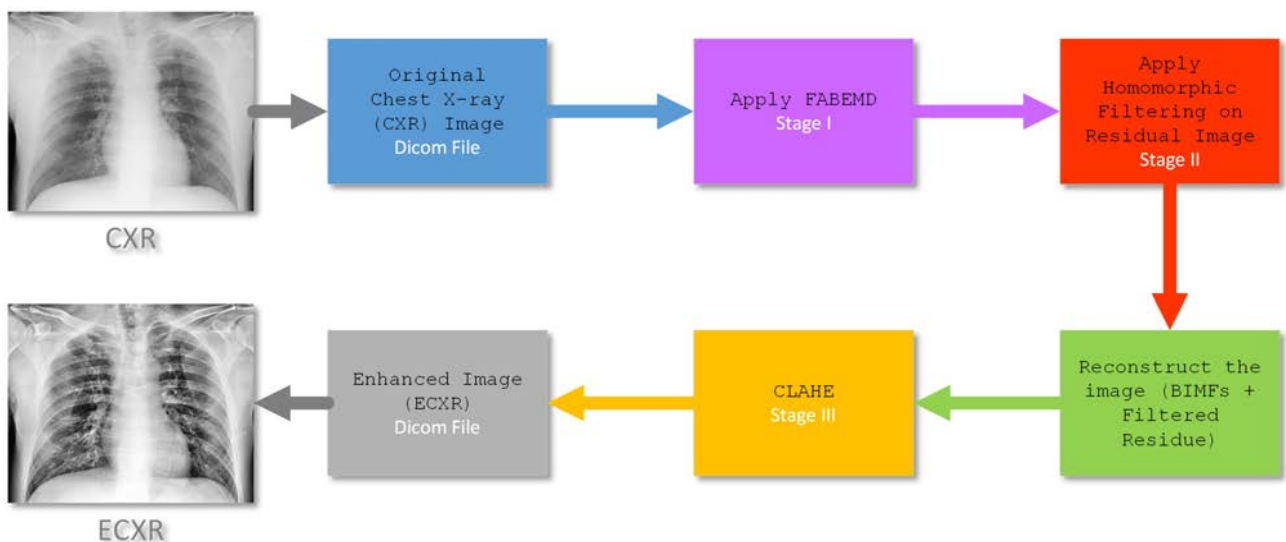

Figure 1. A block diagram of the PACE method as developed in this work. The input CXR is converted into an ECXR through different steps. (I) FABEMD generates the bidimensional intrinsic mode functions (BIMFs) and residual image (BR), (II) the HMF stage is used to filter the BR and correct for brightness

inhomogeneities. Finally, (III) CLAHE is applied on the reconstructed image (BIMFs + filtered BR) to improve the overall contrast and generate the enhanced image (ECXR).

**II. Description and performance evaluation of the algorithm**

*A) Fast and adaptive bidimensional empirical mode decomposition - FABEMD.*

FABEMD [17] is an algorithm, based on a variable window order-statistics filter for the calculation of the intrinsic mode functions (IMFs) in a bi-dimensional space (BIMFs). The filtering process iteratively separates and arranges the different BIMFs depending on the scale and fluctuations, from the faster to the slower ones where the residual image represents the latent information. Compared to other approaches to extract the BIMFs, FABEMD is fast because it does not need a recursive procedure.[18] In this work, the adopted FABEMD algorithm implementation can be summarized by the following steps.

*Step 1 – Local maxima and minima detection*

The $j$-th BIMF $\lambda_j$ of an image $S_i$ is obtained by a neighboring window method [18]. The minima and maxima of $S_i$ represented as a 2D matrix $A = [a_{m,n}]$ ($m$ and $n$ are the row and column respectively) are obtained by window size $w_{ex} \times w_{ex}$ as follows:

$$a_{m,n} = \begin{cases} local\ max\ if\ a_{m,n} > a_{k,l} \\ local\ min\ if\ a_{m,n} < a_{k,l} \end{cases} \quad (1)$$

where [17]:

$$\begin{aligned} k &= m - \frac{w_{ex}-1}{2} : m + \frac{w_{ex}-1}{2}, (k \neq m) \\ l &= n - \frac{w_{ex}-1}{2} : n + \frac{w_{ex}-1}{2}, (l \neq m) \end{aligned} \quad (2)$$

*Step 2 – Size of order-statistic filter window calculation*

At first, we define $d_{adj-\max}$ and $d_{adj-\min}$ to be the adjacent maximum and minimum distance arrays, respectively, which are calculated from each local maximum or minimum point to the nearest nonzero element. Also, $d_{adj-\max}$ and $d_{adj-\min}$ will be sorted in descending order in the array. The gross window width is the computed as follow:

$$\begin{aligned} w_{\max\_en-g} &= \text{minimum}\{d_{adj-\max}\} \\ w_{\max\_en-g} &= \text{maximum}\{d_{adj-\max}\} \\ w_{\min\_en-g} &= \text{minimum}\{d_{adj-\min}\} \\ w_{\min\_en-g} &= \text{maximum}\{d_{adj-\min}\} \end{aligned} \quad (3)$$

*Step 3 – Application of the order statistics and smoothing filters*

Both upper and lower envelopes are computed via the two parameters, $U_{E_j}(x,y)$ and $L_{E_j}(x,y)$, which are defined as:

$$U_{E_j}(x,y) = \max\{\lambda_j(s,t)\} = \frac{1}{w_{s,m} \times w_{s,m}} \sum_{(s,t) \in Z_{x,y}} U_{E_j}(s,t)$$
$$L_{E_j}(x,y) = \min\{\lambda_j(s,t)\} = \frac{1}{w_{s,m} \times w_{s,m}} \sum_{(s,t) \in Z_{x,y}} L_{E_j}(s,t) \qquad (4)$$

where $Z_{x,y}$ is the square region of window size, and $w_{s,m}$ is the window width of the smoothing filter with $w_{s,m}$ equal to $w_{e,n}$. The data operations in Eq. (4) are arithmetic mean filters smoothing local variances, and the average envelope is calculated by using the smoothed envelopes $U_{E_j}$ and $L_{E_j}$. Therefore, the MAX and MIN filters will form a new 2-D matrix for envelope surface which will not change the original 2-D input data.[19]

FABEMD has been used for different purposes in medical imaging,[20] including image retrieval [21] and fusion.[22,23] Here, we have used FABEMD to separate the low varying (latent) from high varying components of CXRs with the aim to correct the luminance contributions by preserving the other information and details. Fig. 2 shows an example of the BIMFs as calculated with the FABEMD approach. By comparing the BIMFs with the original image the more informative BIMFs are the panels (d)-(f). In particular, we note that the edges are well represented on panel (d), the signal is contained in panel (e) while the contrast is mainly enclosed in panel (f). Finally, the residual image in panel (j) represents the latent information which shows evidence of the non-uniform luminance in the image originated by different artifacts as already discussed in the introduction. Similar results have been also observed for the other images elaborated.

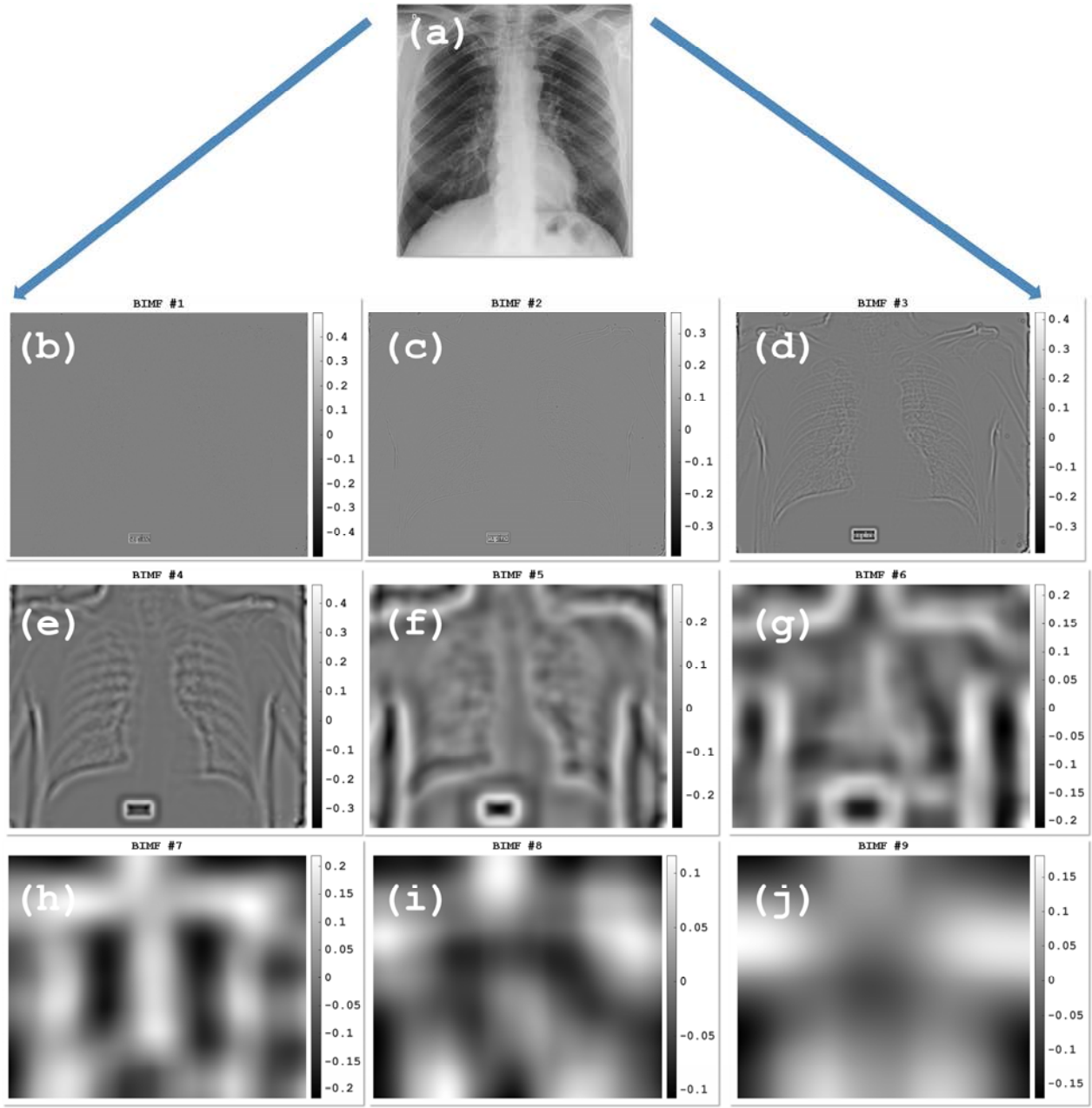

Fig. 2 (a-j): An example of a FABEMD decomposition. (a) Original Image. (b) BIMF #1 and (c) BIMF #2, which contain highly varying components with very low amplitude. (d) BIMF #3, (e) BIMF #4, and (f) BIMF #5 comprehend the more informative components, whereas the (g), BIMF #6, (h) BIMF #7, and (i) BIMF #8 represent the low varying components and (j) the Residue of the calculation.

*B) Homomorphic Filtering - HMF.*

The residual image as computed via FABEMD is then filtered with the HMF $H(x,y)$ to the log-transformed image for correcting the non-uniform illumination of the CXR.[24][25] In particular here, the HMF is designed by using a high-pass filter to reduce or eliminate inhomogeneous effects originating by artifacts in CXR (gray images).[13] Here, we consider a modified version of the high-pass filter $H_{HEF}(x,y)$ based on High-Frequency Emphasis Filter (HEF) given by:

$$H_{HEF}(x, y) = g_L + g_H H(x, y) \tag{5}$$

where $g_L < 1$ and $g_H > 1$, respectively.

The log-transformed image of pixel intensity of an image $I(x, y)$ ($(x, y)$ are the spatial coordinates) can be expressed as a product between reflectance $R(x, y)$ of the object or scene and the intensity of the illumination $L(x, y)$ $I(x, y) = R(x, y) \cdot L(x, y)$.[26] The $L(x, y)$, which represents the image noise, can be reduced by its filtering in the log domain.[27,28] Considering $Q(x, y)$ as the logarithm of the pixel intensity $I(x, y)$:

$$Q(x, y) = \ln(I(x, y)) = \ln(R(x, y) \cdot L(x, y)) = \ln(R(x, y)) + \ln(L(x, y)) \tag{6}$$

In the frequency domain, the Discrete Fourier Transform (DFT) $\Omega$ gives rise to the following expression:

$$\Omega(Q(x, y)) = \Omega(\ln(R(x, y))) + \Omega(\ln(L(x, y))) = \Omega_R(x, y) + \Omega_L(x, y) \tag{7}$$

where $\Omega_R$ and $\Omega_L$ are the DFT of $\ln(R(x, y))$ and $\ln(L(x, y))$ respectively. The filtered image $I'(x, y)$ is computed as $I'(x, y) = \exp(w(x, y))$, $w(x, y)$ being the inverse DFT of the convolution between $Q(x, y)$ and $H_{HEF}(x, y)$ as

$$w(x, y) = \Omega^{-1}(W(x, y)) = \Omega^{-1}(H(x, y) \cdot \Omega_R(x, y)) + \Omega^{-1}(H(x, y) \cdot \Omega_L(x, y)) \tag{8}$$

For the HEF used in this work, we choose $g_L = 0.99$ and $g_H = 1.5$ as computed by using multi-objective optimization (MOO).[29,30] Through the above process, a dynamic range component of the gray-level scale and a contrast enhancement can be obtained.[31] Fig. 3 shows an example of the application of the HMF applied to the Residue achieved with the FABED algorithm as displayed in Fig. 2(j).

In summary, HMF represents the intermediate step in our pipeline and it has been implemented to integrate such filtering process with previous block for image segmentation (FABEMD) and the next reconstruction block which prepares the image for the final contrast enhancement by means of the CLAHE algorithm. The reconstruction block combines together the BIMFs with the filtered residual that will be then processed to generate a final single CXR with improved contrast, preserved details and where the illumination inhomogeneities are dramatically reduced.

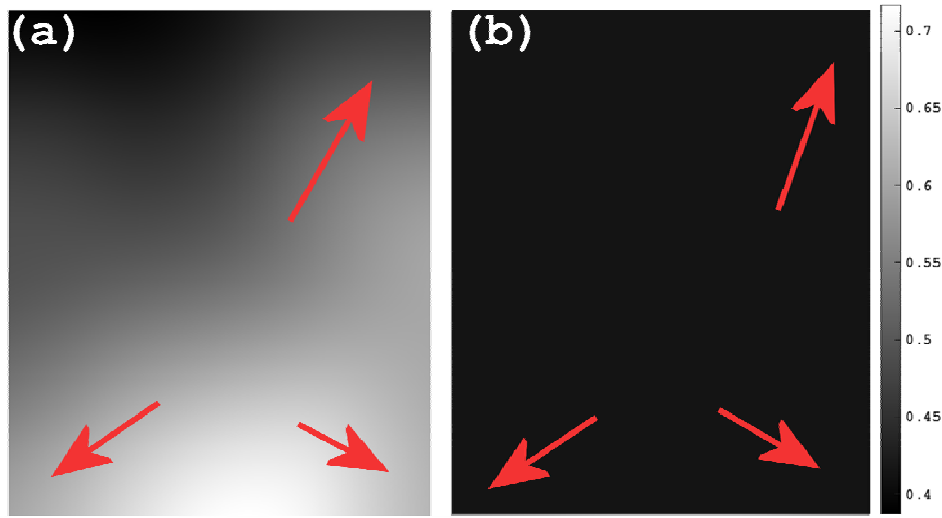

Fig. 3 (a) Residual image as obtained from FABEMD of the same input CXR image as processed in Fig. 2 (panel j). (b) Residual image after being processed using the HMF. As can be shown, the non-uniform luminance of the image has been significantly corrected (see red arrows as reference).

*C) Contrast Limited Adaptive Histogram Equalization - CLAHE*

The key difference between CLAHE and the Adaptive Histogram Equalization [32] is the approach to overcome the noise amplification problem. Basically, CLAHE splits the input image into non-overlapping contextual regions (also called sub-images, tiles or blocks) and then applies histogram equalization to each contextual region, clips the original histogram to a specific value and then redistributes the clipped pixels to each gray level. The size of the contextual region or block size (BS) is typically computed at the point with maximum entropy curvature [33]. The Clip Limit (CL) for the calculations is 0.01 as suggested in Ref. [34], whereas the BS is empirically determined in 16x16 pixels. The redistributed histogram is different from ordinary histogram, because each pixel intensity is limited to a selected maximum. But either the enhanced image and the original image exhibit the same minimum and maximum gray values. In detail, for enhancing the intensity values in each contextual region we have used a uniform distribution, because it has demonstrated to provide better results as suggested in Ref. [35]. Finally, the calculation of the new gray level assignment of the pixels is achieved within a sub-matrix contextual region by using a bi-linear interpolation between four different mappings in order to eliminate boundary artifacts. CLAHE was already applied for the enhancement of contrast in medical images.[25,36–38] Huang proposed CLAHE-DWT using a combination of CLAHE and DWT to overcome the limitations of CLAHE which faces the contrast overstretching and noise enhancement problems. This aspect is central because bright parts of image are unnecessary to be enhanced with respect to the dark ones. The proposed approach motivates the combination of CLAHE with FABEMD and HMF to exploit both direct component segmentation capabilities of the former and correction benefits of the latter as a way to stabilize and preprocess the image before CLAHE is applied.

*D) Metrics definition*

Nowadays, specific processing methods are available [39] to enhance image quality for specific applications.[40,41] In order to assess the improvement, different enhancement metrics can be considered, such as contrast-based metrics computed in spatial domain.[42] Previous works have demonstrated that the combination of several enhancement techniques is required to improve the overall image quality. In this new method, the improvement is focused on homogeneous contrast, visibility, and image detail as they are the most essential and important factors that used in detection, recognition, and monitoring of lung lesions especially for COVID-19 patients. Following the results presented in literature, we have evaluated the performance of the proposed algorithm considering as reference metrics the CII, the ENT and EME,[43] which are examples of such spatial contrast measures.

*Contrast Improvement Index (CII).*

According to the literature, the CII is defined as follows [44]:

$$CII = \frac{C_{processed}}{C_{reference}} \quad (9)$$

where $C_{processed}$ and $C_{reference}$ are the contrast values of the processed and original image, respectively. The contrast $C$ of a region is defined as [42]:

$$C = \sum_{k=0}^{L-1} \sqrt{(k-m)^2 \cdot p(k)} \quad (10)$$

$p(k)$ being the probability to have a specific gray level value .

*Entropy - ENT.*

The entropy,[45] which is a measure of the randomness characterizing the texture of the image, is estimated by the histogram of the image considered as a whole:

$$ENT = -\sum p \times \ln(p) \quad (11)$$

where $p$ is the histogram count for an image segment.

*Measure of Enhancement - EME.*

For an image $x(n,m)$ split into $r \times c$ blocks of size $I_1 \times I_2$, the EME of is defined as [40]:

$$EME_{rc} = \frac{1}{r \times c} \sum_{l=1}^{r} \sum_{k=1}^{c} \left[ 20 \ln \left( CR_{k,l} \right) \right] \quad (12)$$

where $\{k,l\}$ represents the block $B_{k,l}$ considered for the calculations, $EME_{rc}$ depends on the image segmentation into $r \times c$ blocks and the contrast $CR_{k,l}$ (as calculated in the block $B_{k,l}$) is defined as [46]:

$$CR_{k,l} = \frac{I_{max}\{k,l\}}{I_{min}\{k,l\} + c} \quad (13)$$

being $I_{max}$ and $I_{min}$ are the maximum and minimum intensity levels, respectively, for the image $x(n,m)$ inside $B_{k,l}$. The value $c$ is a small constant which is equal to 0.0001 to avoid dividing by 0. The EME measure is suitable for images with attributes like noncomplex segments (e.g. regular geometrical shapes, like for human body parts), small targets in segments, non-periodic pattern in segments, and little to no randomness in segments.[46] In addition, many literatures on contrast enhancement [47–50] adopted EME as an evaluation indicator.

### III. Results and discussions

*A) Evaluation of the algorithm performance*

To demonstrate the performance of the proposed method, 79 baseline CXR images of COVID-19 patients from the University Hospital 'Policlinico G. Martino' in Messina have been analyzed. The original CXR images (3520 × 4280 pixels; 10-bit grayscale; gray-scale value is [0–1024], spatial resolution, 0.175 mm/pixel) are subsampled by 4 to reduce the processing time to 1 minute per single image. Figure 4(a)-(c) show a comparison of the metrics discussed in the previous paragraph as computed with CLAHE (red empty triangles) and the proposed method (blue empty circles) for all the patients. The values are plotted in descending order for a better eyes guide. An improvement for all the images based on all the three metrics have been obtained. In particular, an average increase of 10% in CII, 7.5% in Entropy, and 4.7% in EME have been documented.

A visual example of the image improvement is shown in Fig. 4 (d-f), where an CXR as acquired by the portable X-ray equipment (panel (d)), post-processed with CLAHE (panel (e)) and with the approach proposed here (panel (f)) are displayed, respectively. To better show the improvement, Fig. 4 (g-i) provide a magnified region for evaluating a specific part of the images.

The performance of PACE have been also benchmarked in a public database available (https://github.com/ieee8023/covid-chestxray-dataset)[51] showing, an average increase of 9% in CII, 2.4% in ENT, and 2% in EME.

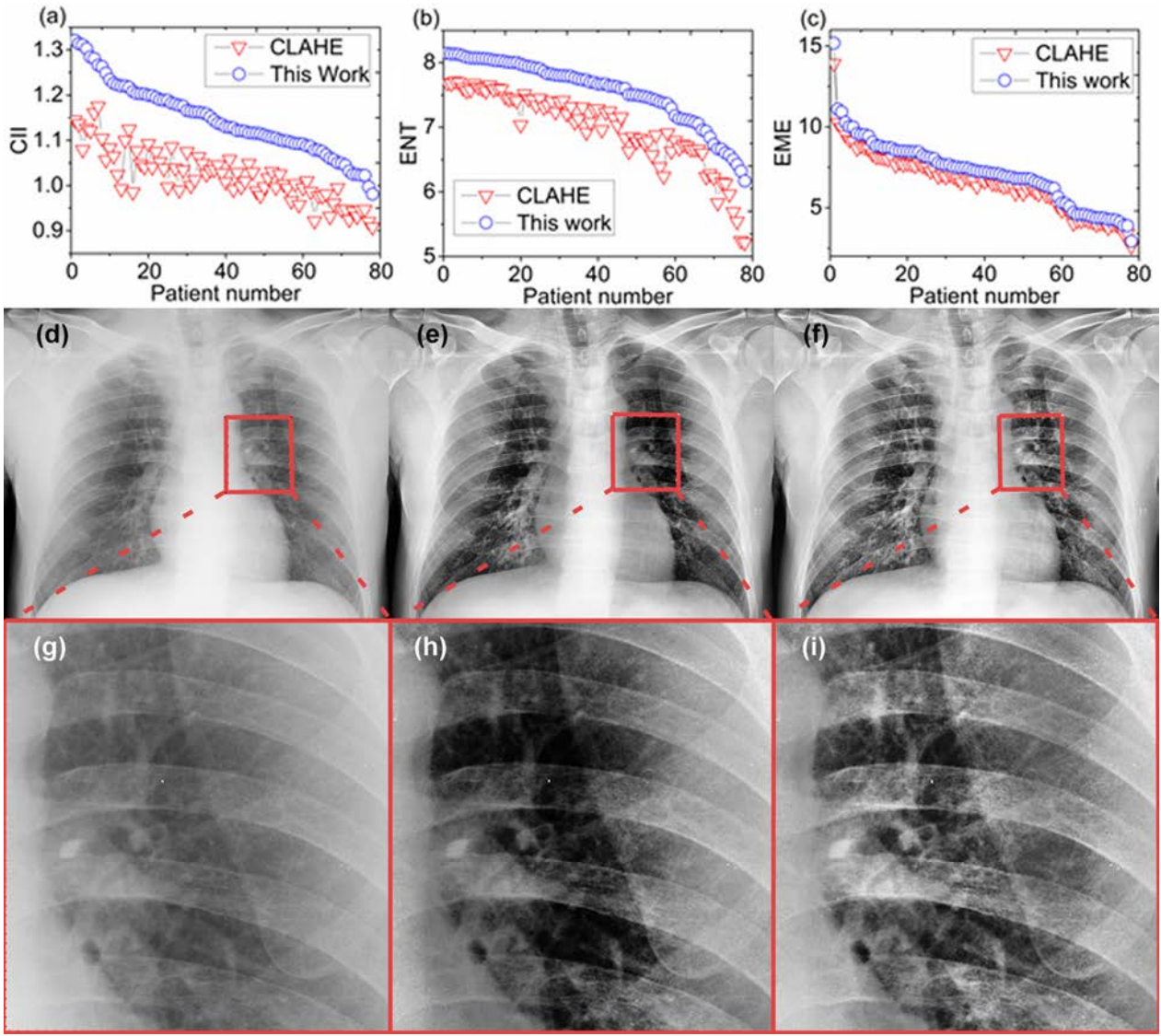

Figure 4 (a-f). The panels (a-c) show a comparison of performance between a literature reference algorithm (CLAHE) (empty red triangles) and the one proposed in this work (empty blue circles) evaluated with three metrics (a) contrast improvement index (CII), (b) entropy (ENT), and (c) and enhancement measurement (EME). The panels (d-f) show an example of the CXR compared. (d) Original image, (e) post processed with CLAHE, and (f) PACE. Finally, the panels (g-i) provide a magnified area with evidence of how lungs are differently shown in original (g), post-processed with CLAHE (h) and with our method (i). PACE enhances both contrast and details in a better way than state-of-the-art reference method.

*B) Clinical evaluation and statistics*

In general, from a clinical point of view, the key achievement is that the number of lesions identified in the original image and the one post processed with CLAHE is the same, while PACE shows the capability to enhance the quality of the CXR enough in order to find more lesions when present. This result is confirmed in 8 patients, for the other cases while the number of lesions is the same for the three images (original, CLAHE, and our method) the velocity of lesions search has been improved as confirmed independently by a third radiologist. To support our claims, in those cases we have also performed the CT scans. Fig. 5 shows a comparison among the original CXR image (panel a), the corresponding CT scan (panel b), and the enhanced CXR (panel c) as computed with PACE. Baseline CXR (a) shows a large peripheral ground-glass pneumonia in the left lung indicated with arrowheads. No lesions can be seen in the right lung. On the other hand, on the enhanced image (c) left lung pneumonia (arrowheads) is better visible as well as further focus of pneumonia in the right lung (arrow). Maximum intensity projection of isotropic coronal reconstruction (b) confirms the presence of both the large left lung lesion (arrowheads) and the small focus of pneumonia in the right lung (arrow). We highlight that on the baseline radiogram it is present (a) overexposition of the right lung due to slight rotation of the patient. On the enhanced image (c) overexposition has been corrected. To analyze how details are preserved and contrast enhanced in Fig. 5 (d-f) show a magnified region for original (d), CT scan (e) and enhanced image (f), respectively. Indeed, a more important aspect is accomplished in 3 cases of the 8 cited above. In those specific cases, it has not been possible to detect any lesions in the original and the CLAHE images, while lesions have been found in the image enhanced by our approach. Such results have been confirmed by the CT scans.

An example for those cases is shown in Fig. 6, where a comparison among original CXR image (a), the corresponding CT scan (b) and the enhanced CXR (c) is displayed. Baseline CXR does not show any lesions. However, both the CT scan and the enhanced image (see arrow) show the presence of a lesion in the left lung. A zoom of the area where this specific lung lesion is located is investigated in panels (d-f). Here, the magnified region is shown from the original image (d), CT scan (e) and with PACE (f), respectively. PACE enables to identify the lesion (black arrow) which is not visible using original CXR. The lesion in confirmed by evaluating the CT scan (white arrow) (e).

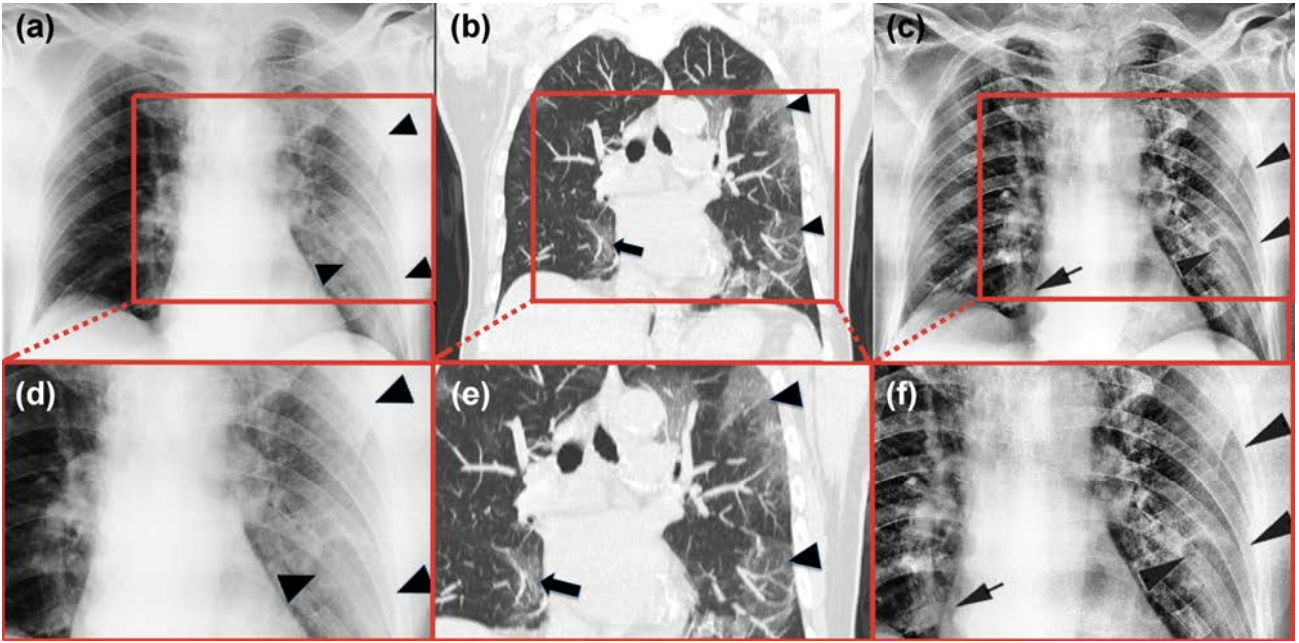

Fig. 5 A comparison among (a) original CXR image, (b) CT image, and (c) enhanced CXR image using PACE where it is possible to show that the number of lesions found in the enhanced image are more than the one identified in the original image. Finally, the panels (d-f) provide a magnified area with evidence of how lungs are differently shown in original (d), CT image (e) and with our method (f).

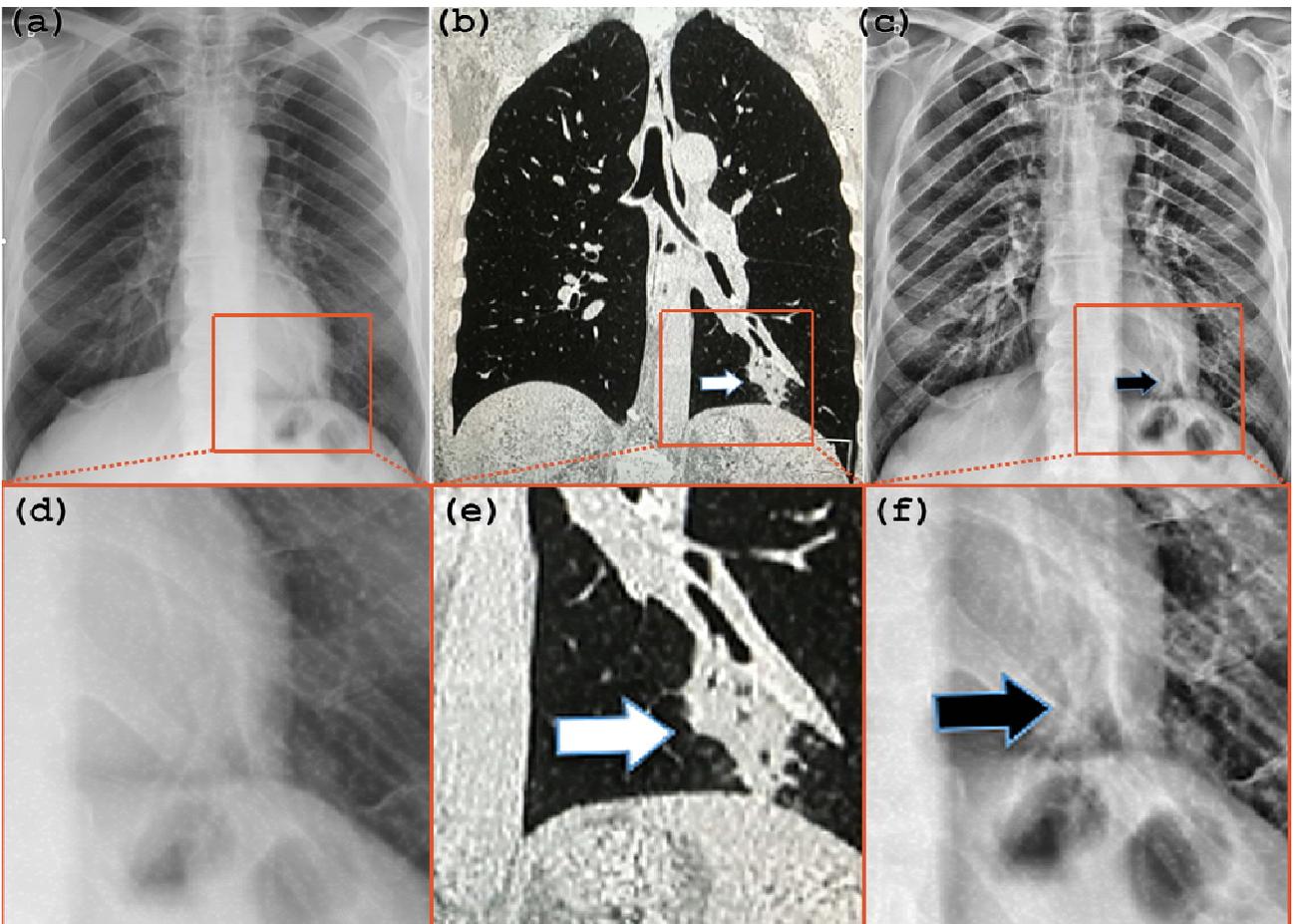

Fig. 6 A comparison among (a) original CXR image, (b) CT image, (c) enhanced CXR image using PACE.

Here, it is possible to see a case where the lesion can be observed in the enhanced image and not in the original one. Lastly, the panels (d-f) provide a magnified area with evidence of how lungs are differently shown in original (d), CT image (e) and with our method (f).

## IV. Summary and Conclusions.

Medical imaging has a significant impact on medical applications, and since the quality of healthcare directly affects the quality of living of a patient, using the images for improving the performance of the medical specialists is an important issue. We have developed an automatic post processing tool, to enhance CXR images for the detection of lungs lesion. These results are generally applicable to image enhancement. On the other hand, the capability to support image monitoring of the healthy status of CODIV-19 patients has been tested in 79 patients. From a clinical point of view, the image enhancement induced by the method implemented in this work originated (*i*) a faster detection of lung lesions in general, (*ii*) in 8 cases it was able to highlight additional lesions, while (*iii*) in 3 out of the 8 cases, lesions have been detected in the enhanced CXR image while baseline CXR did not show any lesions. The clinical findings have been evaluated by two radiologists initially independently each other and a third consultancy has been further used for confirmation. In addition, CT scan verification substantiates the achievement observed in enhanced CXR.

From a technical point of view, the improved performance of PACE approach has been confirmed by three well known metrics, (*i*) contrast improvement index, (*ii*) entropy, and *(iii)* measure of enhancement. On the basis of analysis of these metrics shows that this method preserves the input image details more accurately and gives processed image with better contrast enhancement and reduced brightness inhomogeneities. Concretely, we believe that this tool can support the clinical assistance of COVID-19 patients by enhancing the readability of CXR attained by portable X-ray equipment and in addition furnishing capability to monitor patients in intensive care units.

## Declaration of interest

The authors declare no conflict of interest.

## Ethical statement

The ethical committee at the University Hospital of Messina does not require approval for a work on retrieved and anonymized data. For any information contact the secretary of Ethical Committee at the University Hospital of Messina at the contact information posted in its website (https://pre.polime.it/comitato_etico_interaziendale).

**Author contributions**

G.S. implemented the algorithm and the methodology supported by A.L.C.. G.S, M.G., M.C. and G.F. developed the idea. G.F. coordinated the work. M.G., and G.C. cured the clinical part. G.F. wrote the paper with input from G. S. and M.G. The validation part has been accomplished by G.S., M.G and G.F. All authors commented on the final version of the manuscript.

**Acknowledges**

This work was partially support by PETASPIN association and MARIS s.c.a.r.l.
**Data availability**

The data that support the plots within this paper and other findings of this study are available from the corresponding authors upon reasonable request.

**References**

[1] W. Huda and R.B. Abrahams, Am. J. Roentgenol. **204**, W393 (2015).

[2] WHO, *Coronavirus Disease 2019 (COVID-19) Situation Report-68* (2020).

[3] F. Pan, T. Ye, P. Sun, S. Gui, B. Liang, L. Li, D. Zheng, J. Wang, R.L. Hesketh, L. Yang, and C. Zheng, Radiology 200370 (2020).

[4] F.J. Larke, R.L. Kruger, C.H. Cagnon, M.J. Flynn, M.M. McNitt-Gray, X. Wu, P.F. Judy, and D.D. Cody, Am. J. Roentgenol. **197**, 1165 (2011).

[5] M. Mossa-Basha, C.C. Meltzer, D.C. Kim, M.J. Tuite, K.P. Kolli, and B.S. Tan, Radiology 200988 (2020).

[6] H.Y.F. Wong, H.Y.S. Lam, A.H.-T. Fong, S.T. Leung, T.W.-Y. Chin, C.S.Y. Lo, M.M.-S. Lui, J.C.Y. Lee, K.W.-H. Chiu, T. Chung, E.Y.P. Lee, E.Y.F. Wan, F.N.I. Hung, T.P.W. Lam, M. Kuo, and M.-Y. Ng, Radiology 201160 (2020).

[7] M. Bandirali, L.M. Sconfienza, R. Serra, R. Brembilla, D. Albano, P.F. Ernesto, and C. Messina, Radiology 201102 (2020).

[8] B. Ortiz-Jaramillo, A. Kumcu, L. Platisa, and W. Philips, Signal Process. Image Commun. **62**, 51 (2018).

[9] M.A. Qureshi, A. Beghdadi, and M. Deriche, Signal Process. Image Commun. **58**, 212 (2017).

[10] X. Wang and L. Chen, Signal Process. Image Commun. **58**, 187 (2017).

[11] J. Campbell, M. Pyer, S. Rogers, D. Walter, and R. Reddy, J. Public Health (Bangkok). **36**, 511 (2014).

[12] A.M.Y. Cao, J.P. Choy, L.N. Mohanakrishnan, R.F. Bain, and M.L. van Driel, Cochrane Database Syst. Rev. **12**, online (2013).

[13] A. Walz-Flannigan, D. Magnuson, D. Erickson, and B. Schueler, Am. J. Roentgenol. **198**, 156 (2012).

[14] D. Lv, W. Qi, Y. Li, L. Sun, and Y. Wang, http://arxiv.org/abs/2005.01468 (2020).

[15] E. Tartaglione, C.A. Barbano, C. Berzovini, M. Calandri, and M. Grangetto, http://arxiv.org/abs/2004.05405 (2020).

[16] Y. Oh, S. Park, and J.C. Ye, http://arxiv.org/abs/2004.05758 (2020).

[17] S.M.A. Bhuiyan, R.R. Adhami, and J.F. Khan, in *2008 IEEE Int. Conf. Acoust. Speech Signal Process.* (IEEE, 2008), pp. 1313–1316.

[18] S.M.A. Bhuiyan, R.R. Adhami, and J.F. Khan, EURASIP J. Adv. Signal Process. **2008**, 728356 (2008).

[19] D. Looney and D.P. Mandic, IEEE Trans. Signal Process. **57**, 1626 (2009).

[20] A. Sotiras, C. Davatzikos, and N. Paragios, IEEE Trans. Med. Imaging **32**, 1153 (2013).

[21] W. Liu, W. Xu, and L. Li, in *2007 IEEE 7th Int. Symp. Bioinforma. Bioeng.* (IEEE, 2007), pp. 641–646.

[22] Y.-Z. Zheng and Z. Qin, J. Softw. **20**, 1096 (2010).

[23] A.P. James and B. V. Dasarathy, Inf. Fusion **19**, 4 (2014).

[24] I. Pitas and A.N. Venetsanopoulos, *Nonlinear Digital Filters* (Springer US, Boston, MA, 1990).

[25] H. Wen, W. Qi, and L. Shuang, in *2016 Int. Conf. Robot. Intell. Syst.* (IEEE, 2016), pp. 249–254.

[26] C.-N. Fan and F.-Y. Zhang, Pattern Recognit. Lett. **32**, 1468 (2011).

[27] C.A. Castaño Moraga, C.F. Westin, and J. Ruiz-Alzola, in *Lect. Notes Comput. Sci.* (2003), pp. 990–991.

[28] K. Delac, M. Grgic, and T. Kos, in *Int. Conf. Syst. Signals Image Process.* (2006).

[29] P. Dutta and S. Saha, Comput. Biol. Med. **89**, 31 (2017).

[30] X.-S. Yang, in *Nature-Inspired Optim. Algorithms* (Elsevier, 2014), pp. 197–211.

[31] M.-J. Seow and V.K. Asari, Neurocomputing **69**, 954 (2006).

[32] J.B. Zimmerman, S.M. Pizer, E.V. Staab, J.R. Perry, W. McCartney, and B.C. Brenton, IEEE Trans. Med. Imaging **7**, 304 (1988).

[33] K. Zuiderveld, in *Graph. Gems* (Elsevier, 1994), pp. 474–485.

[34] J. Joseph, J. Sivaraman, R. Periyasamy, and V.R. Simi, Biocybern. Biomed. Eng. **37**, 489 (2017).

[35] Z. Al-Ameen, G. Sulong, A. Rehman, A. Al-Dhelaan, T. Saba, and M. Al-Rodhaan, EURASIP J. Adv. Signal Process. **2015**, 32 (2015).

[36] Sonali, S. Sahu, A.K. Singh, S.P. Ghrera, and M. Elhoseny, Opt. Laser Technol. **110**, 87 (2019).

[37] X. Zhou, Y. Zheng, L. Tan, and J. Zhao, TELKOMNIKA (Telecommunication Comput. Electron. Control. **14**, 1203 (2016).

[38] Y. Tan, G. Li, H. Duan, and C. Li, J. Intell. Syst. **23**, (2014).

[39] K. Panetta, Yicong Zhou, S. Agaian, and Hongwei Jia, IEEE Trans. Inf. Technol. Biomed. **15**, 918 (2011).

[40] S.S. Agaian, B. Silver, and K.A. Panetta, IEEE Trans. Image Process. **16**, 741 (2007).

[41] L. Wu, X. Zhang, and H. Chen, Signal Process. Image Commun. **78**, 254 (2019).

[42] R.C. Gonzalez and R.E. Woods, *Digital Image Processing (3rd Edition)* (2007).

[43] K. Panetta, A. Samani, and S. Agaian, Int. J. Biomed. Imaging **2014**, 1 (2014).

[44] T. Ema, K. Doi, R.M. Nishikawa, Y. Jiang, and J. Papaioannou, Med. Phys. **22**, 161 (1995).

[45] S. Wu, Q. Zhu, Y. Yang, and Y. Xie, in *2013 IEEE Int. Conf. Inf. Autom.* (IEEE, 2013), pp. 521–526.

[46] S. Gupta and R. Porwal, Int. J. Biomed. Imaging **2016**, 1 (2016).

[47] M. Sundaram, K. Ramar, N. Arumugam, and G. Prabin, Appl. Soft Comput. **11**, 5809 (2011).

[48] T. Arici, S. Dikbas, and Y. Altunbasak, IEEE Trans. Image Process. **18**, 1921 (2009).

[49] C. Lee, C. Lee, and C.-S. Kim, IEEE Trans. Image Process. **22**, 5372 (2013).

[50] S.-D. Chen, Digit. Signal Process. **22**, 640 (2012).

[51] J.P. Cohen, P. Morrison, and L. Dao, http://arxiv.org/abs/2003.11597 (2020).